\documentclass[a4paper]{llncs}

\usepackage{fancyvrb}
\usepackage{graphicx}
\usepackage{array}

\begin{document}

\title{Evaluation of a Conversation Management Toolkit for Multi Agent Programming}

\author{David Lillis \and Rem W. Collier \and Howell R. Jordan}

\institute{School of Computer Science and Informatics \\ University College Dublin \\
\email{\{david.lillis,rem.collier\}@ucd.ie\\howell.jordan@lero.ie}
}

\maketitle

\begin{abstract}
The Agent Conversation Reasoning Engine (ACRE) is intended to aid agent developers to improve the management and reliability of agent communication. To evaluate its effectiveness, a problem scenario was created that could be used to compare code written with and without the use of ACRE by groups of test subjects.

This paper describes the requirements that the evaluation scenario was intended to meet and how these motivated the design of the problem. Two experiments were conducted with two separate sets of students and their solutions were analysed using a combination of simple objective metrics and subjective analysis. The analysis suggested that ACRE by default prevents some common problems arising that would limit the reliability and extensibility of conversation-handling code.

As ACRE has to date been integrated only with the Agent Factory multi agent framework, it was necessary to verify that the problems identified are not unique to that platform. Thus a comparison was made with best practice communication code written for the Jason platform, in order to demonstrate the wider applicability of a system such as ACRE.
\end{abstract}

\section{Introduction}

The Agent Conversation Reasoning Engine (ACRE) is a suite of components and tools to aid the developers of agent oriented software systems to handle inter-agent communication in a more structured and reliable manner~\cite{Lillis2011}. To date, ACRE has been integrated into the Agent Factory multi-agent framework~\cite{Muldoon2009} and is available for use with any of the agent programming languages supported by Agent Factory's Common Language Framework~\cite{Russell2011}. Related work is presented in Section~\ref{sec:related}, followed by an overview of ACRE itself in Section~\ref{sec:acre}.

The principal focus of this paper is to describe an experiment that was conducted to evaluate the benefits that ACRE can provide in a communication-heavy Multi Agent System (MAS). Section~\ref{sec:scenario} outlines the motivations underpinning the design of the experiment and discusses how a scenario was designed with these in mind. In particular, we identified a number of variables that should be eliminated so as to ensure that comparisons of ACRE and non-ACRE code would be fair. This scenario was given separately to two classes of  students: one undergraduate and one postgraduate. The students were required to develop agents that could interact with a number of provided agents in order to trade virtual stocks and properties for profit. For each class, the subjects were divided into two groups: one using ACRE and one working without.

The code was analysed both objectively and subjectively and comparisons were made between ACRE and non-ACRE code. This analysis is presented for each of the two classes in Sections~\ref{sec:fudan-experiment} and~\ref{sec:second-experiment} respectively. We then examined some conversation handling code written for the Jason multi agent platform~\cite{Bordini2007} to draw some conclusions about the wider applicability of this work in Section~\ref{sec:jason}. Conclusions and ideas for future work are contained in Section~\ref{sec:conclusions}.

\section{Related Work} \label{sec:related}

Although many well-known agent frameworks and languages have support for some Agent Communication Language (ACL), less attention has been paid to conversations between agents, where two or more agents will exchange multiple messages that are related to the same topic or subject, following a pre-defined protocol. The JADE toolkit provides specific implementations of a number of the FIPA interaction protocols~\cite{Bellifemine2010}. It also provides a Finite State Machine (FSM) behaviour to allow custom interaction protocols to be defined in terms of how they are handled by the agents. Jason includes native support for communicative acts, but does not provide specific tools for the development of agent conversations using interaction protocols~\cite{Bordini2007}. A similar level of support has previously been present within the Agent Factory framework prior to the adoption of ACRE.

Agent toolkits with support for conversations include COOL~\cite{Barbuceanu1995}, Jackal~\cite{Cost1998} and KaOS~\cite{Bradshaw1997}. Other than FSMs, alternative representations for protocols include Coloured Petri Nets~\cite{Huget2003} and Global Session Types~\cite{Ancona2012}.

The comparative evaluation of programming toolkits, paradigms and languages is a matter of some debate within the software engineering community. One popular approach is to divide subjects into two groups with each asked to perform the same task~\cite{Hochstein2008,Luff2009,Rossbach2010}. To the greatest extent possible, objective quantitative measures are used to draw comparisons between the two groups. A common concept to evaluate is that of \textit{programmer effort}, which has been measured in numerous different ways including development time~\cite{Hochstein2008,Luff2009}, non-comment line count~\cite{Luff2009} and non-commented source code statements~\cite{VanderWiel1997}. These measures are used to ensure that a new approach does not result in a greater workload being placed on developers using it.

\section{ACRE} \label{sec:acre}

ACRE is a framework that aims to give agent programmers greater control over the management of conversations within their MASs. It is motivated by the fact that many widely-used Agent Oriented Programming (AOP) languages and frameworks require communication to be handled on a message-by-message basis, with no explicit link between related messages. It frequently tends to be left as an exercise for the developer to ensure that messages that form part of the same conversation\footnote{In this work we draw a distinction between \textit{protocols}, which define how a series of related messages should be structured and \textit{conversations}, which are individual instances of agents following a protocol.} are interpreted as such. This section provides a brief overview of ACRE's capabilities. For further information, it is presented in some detail in~\cite{Lillis2011}.

The principal components of ACRE are as follows:
\begin{itemize}
	\item \textit{Protocol Manager (PM):} The PM is shared amongst all agents residing on the same agent platform. It accesses online protocol repositories and downloads protocol definitions at the request of agents.
	\item \textit{Conversation Manager (CM):} Each agent has its own CM, which is responsible for monitoring all its communication so as to group individual messages into conversations. Messages are compared to known protocol definitions and existing conversations to ensure that they are consistent with the available descriptions of how communication should be structured.
	\item \textit{ACRE/Agent Interface (AAI):} Unlike the PM and CM, the AAI is not platform-independent, with its implementation dependent on the framework and/or AOP language being employed. The AAI serves as the API to the CM and PM: agents can perceive and act upon the ACRE components.
\end{itemize}

Additionally, the wider ACRE ecosystem provides an XML format for defining custom interaction protocols, a standard for the organisation of online protocol repositories and a suite of tools to aid developers in communication handling. These tools include a graphical protocol designer, a runtime conversation debugger, a GUI to manage and explore protocol repositories and a runtime protocol view to show what protocols have been loaded on an agent platform.

\begin{figure}[!htb]
\centering
\includegraphics[width=8cm]{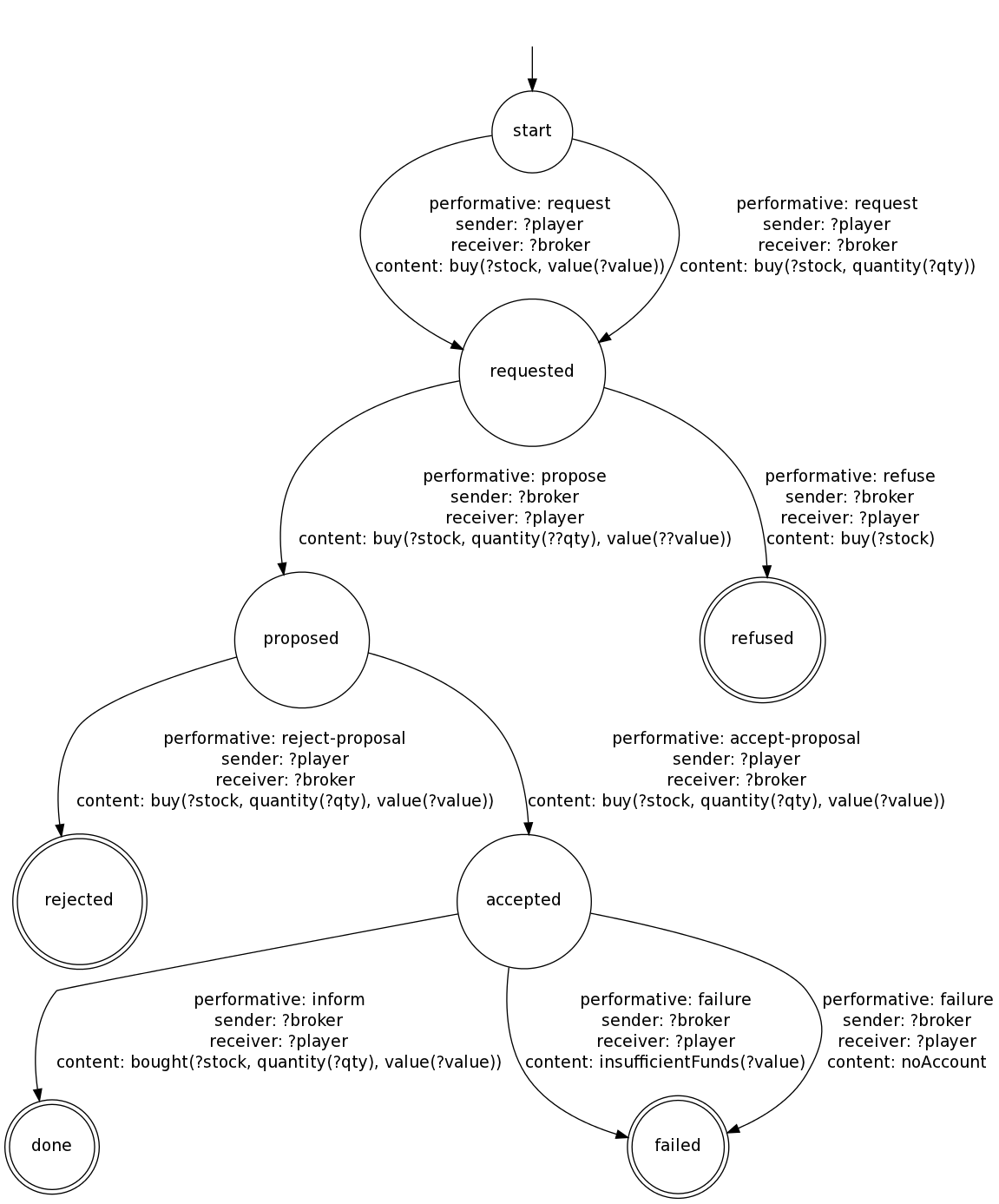}
\caption{An ACRE protocol}
\label{fig:broker-buy}
\end{figure}

ACRE's representation of protocols uses FSMs where transitions between states are activated by the sending and receiving of messages. An example of this can be seen in Figure~\ref{fig:broker-buy}. Here, the conversation is begun by a message being sent that matches any of the transitions emanating from the initial ``Start'' state. When this occurs, the variables (prefixed by the \texttt{?} character) are bound to the values contained in the message itself. This will, for example, fix the name of the conversation initiator (bound to the \texttt{?player} variable) and the other participant (\texttt{?broker}) so that subsequent messages must be sent to and from those agents.

If a message fails to match the specification of the relevant protocol, the CM will raise an error to make the agent aware of this. This feature is not readily available in existing AOP languages, as these typically depend on event triggers to positively match an expected message, with unrecognised or malformed messages being silently ignored if they match no rule.

From an AOP developer's point-of-view, ACRE facilitates the implementation of interaction protocols by automatically checking messages against known protocols and conversations, providing information about the state of conversations as well as making available a set of actions that operate on them. The information available includes the participants in a conversation, the conversation state, the messages that make up the conversation history and the protocol a conversation follows. Additionally, the agent is made aware of conversation events such as the conversation being advanced, cancellation, timeouts and errors (such as messages failing to match a transition in a defined protocol). Available actions include creating or advancing a conversation, cancelling existing conversations or communicating protocol errors to other participants.

\begin{figure}[!htb]

 	\begin{Verbatim}[fontsize=\footnotesize]
+initialized : true <-
  acre.init,
  acre.addContact(agentID(banker,addresses("local:localhost"))),
  acre.start(open,banker,request,openAccount);
 	
+conversationAdvanced(?cid,done,?length) :
conversationProtocolName(?cid,open) <-
  +bankAccount;
 	\end{Verbatim}
 \caption{Communicating in AF-AgentSpeak using ACRE.}
 \label{fig:ACRE-code}

\begin{Verbatim}[fontsize=\footnotesize]
+initialized : true <-
+openingAccount(banker),
.send(request,agentID(banker,addresses("local:localhost")),openAccount);

+message(inform,agentID(?sender,?addr), openedAccount(?id,?amt)) :
openingAccount(?sender) <-
+bankAccount;
\end{Verbatim}
	\caption{Communicating in AF-AgentSpeak without using ACRE.}
	\label{fig:non-ACRE-code}
\end{figure}

Figures~\ref{fig:ACRE-code} and~\ref{fig:non-ACRE-code} show short code samples, written in AF-AgentSpeak (an adaptation of Rao's AgentSpeak(L)~\cite{Rao1996} that is inspired by Jason~\cite{Bordini2007}), showing how a simple interaction could be carried out with and without ACRE respectively. From these samples, the difference in approach may be seen.

The ACRE code adds the contact details of another agent only once, referring thereafter only to its name (``banker'' in this example). Additionally, the events, beliefs and actions refer to details of a conversation (identified by a unique conversation identifier, \texttt{?cid}). The second rule reacts to a \texttt{conversationAdvanced} event, which indicates that a conversation has been advanced by the communication of a message to a state named ``done''. This state name is taken from the definition of the protocol that the conversation is following. The variables \texttt{?cid} and \texttt{?length} will be matched to the unique identifier of the conversation and the length of the conversation respectively. The conversation length is included to ensure that different messages that result in the same state (where the protocol contains a loop) will produce distinct events.

 Without ACRE, messages are dealt with individually, with conversation handling left to the developer (in this example, the \texttt{openingAccount(banker)} belief records that the Player is engaging with the Banker agent to open a bank account). In addition, the name and address of the other agent must be provided each time a message is sent, and matched for every incoming message.

\section{Evaluation Experiment} \label{sec:scenario}

In order to evaluate the usefulness of ACRE, an experiment was designed whereby two groups of participants would write AOP code to tackle a particular problem. One group was required to write their code using the capabilities of ACRE whereas the other worked without ACRE. Before describing the experiment that was conducted, it is important to outline the motivations involved in its creation.

\subsection{Motivations}

The design of the problem was guided by the desire for the scenario to be \textit{communication-focused}, \textit{accessible} and \textit{reproducible}, with a \textit{clearly defined implementation sequence} and a clear \textit{reward for active agents}, so that is it not possible for an agent to benefit by being inactive. Some of these motivations are specific to this particular experiment but many are desirable properties of any experiment where programming languages, paradigms or tools are being evaluated comparatively. These motivations are discussed in more detail below.

\begin{itemize}
   \item \textit{Focus on Communication:} The aim of the evaluation was to engage in a scenario that was communication-heavy. To facilitate this, it was decided that the problem should require developers to create a single agent, so that they would not be distracted from the core focus by having to deal with issues such as co-ordination and co-operation. A group of ``core agents'' would be provided with accompanying protocols, with which the participant's agent must interact. Communicating with the core agents should be required from the beginning, with no progress being possible without communication.
   \item \textit{Accessibility:} As the primary focus of the scenario is communication, little complex reasoning should be required to build a basic implementation that can perform well.
   \item \textit{Reproducibility:} Every non-deterministic environment state change and core agent decision should be recorded, so that the experiment can be exactly replicated. Given a deterministic player agent, each replication should yield identical results.
   \item \textit{Clearly defined goal:} From the point of view of participants, the assigned tasks, time allowed and scoring criteria should be clear.
   \item \textit{Clearly defined implementation sequence:} Participants should not be able to gain a better score merely by implementing parts of their solution in a different order. The easiest way to ensure this is to fix the task order. In the context of a communication-heavy experiment, task ordering may be enforced by making later protocols dependent on others, thus avoiding the need to monitor participants directly.
   \item \textit{Rewards for Active Agents:} There should be no features of the experiment where an agent implementing that feature is at a disadvantage when compared to an agent that does not implement it. 
\end{itemize}

\subsection{Scenario}

The scenario chosen for the experiment was a simple asset trading game. Each participant was asked to develop an agent named \textit{Player}. This agent was required to interact with a number of provided core agents in order to buy and sell virtual stocks and properties so as to increase the amount of money they had. The core agents were as follows:

\begin{itemize}
   \item \textit{Banker:} The Banker agent maintains a Player's bank account. The first task of each agent is to interact with the Banker to open an account, in which an initial amount of virtual currency is placed.
   \item \textit{Stockbroker:} This agent is responsible for buying and selling stocks. Players earn money by buying from the Stockbroker and selling later at a profit.
   \item \textit{Guru:} The Guru agent is aware of how the market operates and can provide tips on which stocks will rise quickly in price and which should be avoided.
   \item \textit{Auctioneer:} Properties can be bought from the Auctioneer agent. The value of properties rises quickly so they offer a method of making greater profits than on the stock market alone.
   \item \textit{Bidder:} Bidders will participate in auctions organised by the Player to buy properties that the Player has previously purchased from the Auctioneer.
\end{itemize}

A number of features of the game were created with the motivations outlined above in mind. The \textit{focus on communication} is maintained by providing a number of protocols that the core agents are capable of following. Each protocol is based on one or more of the FIPA interaction protocols. These are summarised in Table~\ref{tab:protocols}. An illustration of one of these protocols is shown in Figure~\ref{fig:broker-buy}, namely the protocol invoked to buy stock from the Stockbroker agent. Similar illustrations were available for all protocols.

\begin{table}
\caption{Core Agent Protocols} \label{tab:protocols}
\centering
\begin{tabular}{|l|l|l|l|}
\hline
\textbf{Agent} & \textbf{Protocol} & \textbf{Based On} & \textbf{Purpose} \\
\hline
Banker 			& open		& request & Open a bank account \\
				& enquiry	& query & Query your bank balance \\
\hline
Stock Broker	& listing 	& query & Get a list of available stocks \\
				& price		& query & Query the price of a particular stock \\
				& portfolio	& query & Query details of stocks currently owned \\
				& buy		& request & Buy a quantity/value of a particular stock \\
				& sell		& request & Sell a quantity/value of a particular stock \\
\hline
Guru			& subscribe	& subscribe & Subscribe to the guru agent's stock tips \\
\hline
Auctioneer		& subscribe	& subscribe, & Subscribe to details of new auctions and \\ 
				&			&  english-auction & participate in these auctions. \\
\hline
Bidder			& sell		& contract-net & Sell a property \\
\hline
\end{tabular}
\end{table}

The \textit{clearly defined goal} of the task is maximise capital. Thus participants are aware of what they need to do in order to be successful.

As a result of the \textit{accessibility} motivation, the scenario is designed so that a Player can be successful using a simple strategy, without requiring advanced reasoning. This is important when experiment participants are time-constrained.

By default, the movement of stock prices is determined randomly. An internal clock is used to track the time of the experiment: stock prices may change on every ``tick'' of the clock. A \textit{reproducible} experiment may be conducted by loading pre-prepared stock prices at the beginning of the experiment, thus ensuring that the price movements are repeated across multiple experiments.

The ordering of the core agents in the list above reflects the order in which they should be interacted with, thus creating a \textit{clearly defined implementation sequence}. The tasks are designed so that successfully completing a task will be dependent on previous tasks being completed first. Although there are no technical restrictions on the order in which participants may choose to implement the tasks, these dependencies discourage them from writing their implementations in a different order. For example, selling items to bidders is impossible before interacting with the Auctioneer to buy properties. However, these cannot be bought prior to interacting with the stock market, as the minimum property price is deliberately set to be higher than the Player's initial capital. Similarly, the advice of the Guru is useless unless an agent can use it when interacting with the Stockbroker, and it is impossible to buy or sell stocks without having previously opened an account with the banker. 

To \textit{reward active agents}, the stock price calculation mechanism is intentionally artificial, in that the price of stocks always rises. This rewards developers for implementing features, to the detriment of idle agents. If stock prices can fall, an agent that does not participate in the market may end with more money than one that has interacted with a Stockbroker but that has lost money in doing so.

\section{First Experiment} \label{sec:fudan-experiment}

The experiment was first conducted with a final year undergraduate class in Fudan University, Shanghai, China. The evaluation was conducted as part of a module on Agent Oriented Programming. None of the participants had previous experience in developing MASs or in using an AOP language. 

For consistency, all participants were required to write their code in AF-AgentSpeak so as to run within the Agent Factory framework. This removes any bias associated with the use of different AOP languages or frameworks.

Students were allowed three hours in a supervised laboratory setting in which to create their solutions. The fixed time period allows quantitative comparisons to be done with regard to the number of protocols each student implemented. The choice of a supervised in-class test ensured that each student submitted their own work. Subjects were permitted to access lecture notes and refer to manuals and user guides relating to Agent Factory, AF-AgentSpeak and ACRE.

The participants were divided into two groups of equal size using a random assignment: one group was requested to implement their solution using the extensions provided by ACRE whereas the other group was requested to implement their solution using the existing Agent Factory message-passing capabilities. In preparation for the experiment, a practical session was conducted so the participants had the opportunity to gain familiarity with both forms of message handling. This occurred a week prior to the evaluation so as to afford the students sufficient time to get accustomed to agent communication. Previous practical sessions held as part of the module exposed the participants to other aspects of AF-AgentSpeak and AOP programming in general.

The class consisted of 46 students in total, therefore 23 were asked to implement each type of solution. One student from the non-ACRE group did not attend the evaluation, leaving 45 submissions. Additionally one other student from the non-ACRE group instead submitted a solution that did use ACRE.

Of the remaining 21 students in the non-ACRE group, one submission was not included in this research as only one protocol had been attempted and this attempt had not been successful, leaving a total of 20 non-ACRE submissions.

24 submissions were received that had used ACRE. Again, one of these has not been included in this analysis, as the agent submitted did not successfully interact with any core agent. This left a total of 23 submissions using ACRE.

The submissions were evaluated using both objective and subjective measures. Initially, some simple objective measures were employed to measure programmer effort (following the examples outlined in Section~\ref{sec:related}). Following this, the implementations were examined to identify any issues that the implementations may have failed to address.

\subsection{Objective Measures} \label{sec:fudan-objective}

The principal aim of ACRE is to help developers to deal with complex communication more easily. As such, it is important to ensure that the use of ACRE does not add to the effort required to develop MASs. Objective measures are required to attempt to quantify programmer effort. Two simple metrics were employed for this purpose: 1) the number of protocols implemented within a specified time period and 2) the number of non-comment lines of code per protocol.

The first of these can be used to compare the two participant groups in terms of the time taken to implement protocols. Because of the ordering of the tasks, participants are encouraged to implement the protocols in the same order as one another. For example, it is not productive for a participant to begin by implementing the complex auctioneer protocols while others are implementing the simple protocol required to open a bank account. This helps to prevent the metric being skewed by certain subjects being delayed by the order in which they chose to implement their system.

Because there is variation in the number of protocols successfully implemented by each participant, the count of code lines is averaged for the number of protocols implemented. Again, the clear implementation sequence means that this is to the greatest extent possible measuring like-with-like.

\begin{table}[!hbt]
\centering
\caption{Objective measures of programmer effort for the first experiment.} \label{tab:speed}
\begin{tabular}{|l|c|c|}
\hline
 & Protocols Implemented & Lines per Protocol  \\
\hline
 ACRE & 5.43 & 18.35 \\
 Non-ACRE & 5.85 & 27.06 \\
\hline
\end{tabular}
\end{table}

Table~\ref{tab:speed} shows the average number of protocols implemented by each group, along with the average number of non-comment lines of code present per protocol implemented. For the number of protocols implemented, the difference is not statistically significant using an unpaired \textit{t}-test for $p=0.05$. There is, however, a statistically significant difference in lines per protocol using the same test.

It can be seen that participants in the ACRE group implemented marginally fewer protocols on average within the three-hour period. However, as this difference is not significant, it indicates that the speed of development is comparable whether ACRE is being used or not. This suggests that ACRE does not impose a steep learning curve compared to traditional methods of conversation handling.

It is interesting that the ACRE implementations did tend to have significantly fewer lines of code per protocol. Although this is a somewhat crude metric, it does suggest that the automated conversation handling of ACRE may reduce the amount of code that is required in order to successfully implement protocols.

Although objective measures are desirable in any evaluation, it is important to take a subjective view of the code also. While no significant difference in the amount of programmer effort required was observed, these metrics do not capture the quality of the implementations. The next section presents subjective analysis of the code submitted that attempts to identify this quality.

\subsection{Subjective Assessment} \label{sec:fudan-subjective}

The adoption of ACRE will be most beneficial if it improves code quality. To gauge this, the code was examined and a number of issues were found to be prevalent in the non-ACRE code. These issues meant that the solutions that were submitted were very closely tied to the scenario as presented, and would have required much more additional work to be done for an extended MAS.

By its nature, AF-AgentSpeak reacts to the receipt of messages using rules that have a \textit{triggering event} and a \textit{context}. The triggering event is a some event that has occurred (i.e. a change in the belief base) whereas the context is a set of beliefs that should be present for the rule to be relevant. When writing non-ACRE code in AF-AgentSpeak, the event that triggers the rule is the receipt of an incoming message and the context consists of beliefs about the state of the conversation (checking the sender, ensuring that the message follows the correct preceding message etc.). This can be seen in Figure~\ref{fig:non-ACRE-code}. For the issues that were identified, the context of the rules were not written in the best way possible.

\subsubsection{Identification of Issues} \label{sec:fudan-issues}

The particular issues that were identified are as follows. Words in parentheses are short descriptions that are used to refer to each issue in the ensuing analysis:
\begin{itemize}
	\item \textit{No Checking of Message Senders (Sender):} Incoming messages were matched only using their performative and content, with no checks in place to ensure they were sent by the correct agent.
	\item \textit{No Checking of Conversation Progress (Progress):} As the messages followed particular protocols, they were exchanged in a clearly-defined sequence. Many programmers did not attempt to check the context of messages that were received, treating them as individual messages.
	\item \textit{Hard-coded name checking (Name):} When the message sender was checked, it was frequently the case that the name of the sender was hard-coded.
	\item \textit{Checking addresses (Address):} Agent's addresses were also hard-coded.
\end{itemize}

While a solution that includes these issues is capable of successfully participating in the trading scenario, their presence means that additional effort would be required to adapt the solution for a more open agent system or if the scenario were to be extended with additional protocols and/or agents.

When an agent fails to check the identity of the sender, this may have unintended consequences. For example, a Player agent would normally react to a recommendation from the Guru to buy a particular stock by following that recommendation. In a more open MAS, a malicious agent may send recommendations that cause other players to buy stocks that will not perform well. All that would be required is to send a message with the same performative and content as the Player would expect from the Guru.

Similarly, a Player that does not record the state of conversations is also susceptible to exploitation. For example, the protocol for buying stock (shown in Figure~\ref{fig:broker-buy}) insists that the Player must have accepted a proposal to buy stock before the purchase proceeds. However, without a notion of conversation, a Player may be persuaded that it has bought some quantity of a stock without it being involved in the process. If this is combined with the message sender not being checked, an agent other than a Stockbroker may trick a Player into buying stocks it had no intention to buy. As an aside, it is interesting to note that those agents that did record the progress of a conversation tended to use the state names provided in the ACRE FSM diagrams, suggesting that even for developers who do not use ACRE (or lack ACRE support in their AOP platform of choice), the availability of protocols defined in this way is useful for visualising and reasoning about conversations.



The two other issues relate to the difficulty in re-using the code for an extended scenario where conversations are conducted with different agents and/or multiple platforms. Even where the sender of a message was checked, it was frequently the case that this was hard-coded into the context of every rule. As such, the code was capable of conducting a conversation only with an agent of a specific name. If the scenario were to be extended so that multiple agents were capable of engaging in the same protocols (e.g. two Stockbroker agents that handled a different set of stocks) then these rules would all require re-writing to allow for additional agents. Similarly, hard-coding the addresses in the context of each rule limits the code to single platform MASs. Adding agents on another platform will also require all rules to be rewritten.

Because ACRE's Conversation Manager automatically performs checking of this type, such issues cannot arise within ACRE code. As illustrated in the code from Figure~\ref{fig:ACRE-code}, the triggering event is typically that a conversation has advanced, with the context being used to check other details about the conversation. Conversation participants need only be named when the conversation is initiated.

Assuming a pre-existing conversation, a \texttt{conversationAdvanced} event can only be triggered by a message that has been sent by the existing participant to the conversation and has a performative and content that match the next expected message in the protocol. If a message is sent by a different agent, an \texttt{unmatchedMessage} event is raised to indicate that the message does not belong to any particular conversation. This means that it is not necessary to check the message sender for each rule relating to communication, as the event cannot be triggered by the wrong agent.\footnote{ACRE does not protect against messages sent by an agent other than that identified in the message's \texttt{sender} field. This type of secure communication is considered to be a task for the underlying Message Transport Service.}

This automatic checking also guards against out-of-sequence messages. In the example of the buying protocol shown in Figure~\ref{fig:broker-buy}, the Conversation Manager insists that the messages be communicated in the specified sequence, so the Stockbroker cannot inform the Player of a successful purchase unless the Player has previously agreed to that purchase (again, an \texttt{unmatchedMessage} event would be triggered).

Thus the use of ACRE automatically protects against these issues, meaning that were the MAS be more open or the scenario extended, far less effort would be required to adapt the existing ACRE code to the altered circumstances. ACRE can be seen to prevent certain coding styles that would restrict the extensibility and reusability of communication-handling code. Although subjects were not explicitly instructed to create generic code with wider applicability, we believe that ACRE's prevention, by default, of these type of problems is a strong argument in its favour.

\subsubsection{Prevalence of Issues} \label{sec:fudan-prevalence}

For each of the issues outlined in the previous section, it is necessary to measure how prevalent they are amongst the implementations that were submitted. Each implementation was given one of three classifications with regard to each of the four issues identified:
\begin{itemize}
	\item \textit{Not susceptible:} The issue was not present for any rule in the implementation.
	\item \textit{Totally susceptible:} The issue in question was present in every rule where it was relevant.
	\item \textit{Somewhat susceptible:} The issue was present for some relevant rules but not all. This ranges from those implementations where a check was omitted only once to those where the check was only performed for one rule.
\end{itemize}

In relation to hard-coded name and address checking, these issues could not be present in agents that were totally susceptible to the issue of checking message senders. For those agents that did not check message senders at all, it is not possible for these other issues to arise. For this reason, in the following analysis, the figures shown for these two issues are displayed as a percentage of those agents in which is was possible for them to arise.

\begin{table}[!htb]
	\centering
	\caption{Issues present in non-ACRE code for the first experiment.}
	\label{tab:fudan-issues}
	\setlength{\extrarowheight}{5pt}
	\begin{tabular}{|l|l|l|l|l|}
\hline

& \textbf{Sender} & \textbf{Progress} & \textbf{Name} & \textbf{Address} \\
\hline
	Totally Susceptible 	& 	40\% (8) & 	55\% (11) & 	67\% 	(8) & 	25\% (3)\\
	Somewhat Susceptible 	& 	30\% (6) & 	30\% (6) & 	0\% (0)	& 	17\% (2) \\
	Not Susceptible 		& 	30\% (6) & 	15\% (3) & 	33\% 	(4) & 	58\% (7) \\
	\hline
	\end{tabular}
\end{table}

Table~\ref{tab:fudan-issues} shows the prevalence of the issues amongst the non-ACRE submissions. Figures in parentheses are the absolute number of subjects each percentage relates to. Agents that were totally susceptible to the \textit{Sender} issue are not included in the calculations for \textit{Name} or \textit{Address}, as these are only based on the number of agents that attempted to check the message sender.

From these, it can be seen that the issues raised were widespread amongst non-ACRE developers. The hard-coding of agents' addresses is the only issue that was found in less than half of relevant agents.

Over two thirds of agents would react to messages sent by the wrong agent at least some of the time, with 40\% failing to ever check the identity of a message sender. Of those that did check, two thirds hard-coded the name of the sender, which would require every rule to be re-written if the scenario was to be altered.

Just three participants (15\%) always checked that messages were in the correct order. On further analysis, all of these implementations were somewhat susceptible to the Sender issue, which meant that there were no submissions where no issues were found.

The following Section describes a second run of the same experiment using different participants. The results of this can be compared to those presented above to further demonstrate the extent to which the issues identified here appear in code written by more experienced programmers.

\section{Second Experiment} \label{sec:second-experiment}

The same experiment was repeated later with a different set of participants. On this occasion, the subjects were part of a part-time Masters-level Agent Oriented Software Engineering module in University College Dublin, Ireland. These students are experienced professional software developers working in industry, although none had prior experience with AOP.

For this group, classes were conducted daily for five days. This included practical work each afternoon to allow students to become familiar with AOP. Communication handling and ACRE were introduced on the fourth day and the evaluation occurred on the fifth day. Thus these students had less preparation time than those in the first experiment.

Students were again divided into two groups by random assignment. In a class of 19 students, 10 submitted ACRE-based solutions while the remaining 9 students created non-ACRE agents. The scenario was conducted in exactly the same way as for the first experiment.

\subsection{Objective Measures}
The objective metrics employed were the same as in Section~\ref{sec:fudan-objective}. The results are presented in Table~\ref{tab:ucd-objective}. As with the first experiment, the ACRE participants implemented marginally fewer protocols, though this difference was not statistically significant. Overall, fewer protocols were implemented compared to the first experiment. This may have been as a result of the shorter preparation time available to these students. For the lines of code per protocol metric, it can be seen that both groups of ACRE students wrote a very similar amount of code. However, it is interesting to note that the number of lines written per non-ACRE protocol in the second experiment is lower. Again, however, this difference is not statistically significant.

\begin{table}[!hbt]
	\centering
	\caption{Objective measures of programmer effort for the second experiment.}
	\label{tab:ucd-objective}
	\begin{tabular}{|l|c|c|}
		\hline
		& Protocols Implemented & Lines per Protocol  \\
		\hline
		ACRE & 4.4 & 18.93 \\
		Non-ACRE & 4.67 & 14.87 \\
		\hline
	\end{tabular}
\end{table}

Given the small sample size in the second experiment, it is difficult to draw concrete conclusions based on quantitative objective analysis. However, the metrics do add support to the argument that ACRE does not add to the amount of programmer effort required to create a communication-heavy agent program.

\subsection{Subjective Analysis} \label{sec:ucd-subjective}

For a subjective analysis of the second set of students' submitted code, it is interesting to determine the extent to which the issues that became apparent in the first experiment are also present in the second. Table~\ref{tab:ucd-issues} shows the prevalence of these issues in the non-ACRE code from the second experiment. The percentages in Table~\ref{tab:ucd-issues} are of the total of 9 non-ACRE students for the \textit{Sender} and \textit{Progress} issues. For the \textit{Name} and \textit{Address} issues, these are calculated from the 4 participants that were either totally or somewhat susceptible to the \textit{Sender} issue. This is calculated in the same was as in Table~\ref{tab:fudan-issues}.

\begin{table}[!htb]
	\centering
	\caption{Issues present in non-ACRE code for the second experiment.}
	\label{tab:ucd-issues}
	\setlength{\extrarowheight}{5pt}
	\begin{tabular}{|l|l|l|l|l|}
		\hline
		
		& \textbf{Sender} & \textbf{Progress} & \textbf{Name} & \textbf{Address} \\
		\hline
		Totally Susceptible 	& 	56\% (5) & 78\% (7) & 100\% (4)	& 	0\% (0)\\
		Somewhat Susceptible 	& 	0\% (0)	 & 22\% (2) & 0\% (0)	& 	0\% (0) \\
		Not Susceptible 		& 	44\% (4) & 0\% (0)	& 0\% (0) 	& 	100\% (4) \\
		\hline
	\end{tabular}
\end{table}

A notable difference between these results and those arising from the first experiment is that on this occasion, no student hard-coded addresses when checking message senders. It is of interest, however, that several students hard-coded addresses for some outgoing messages. This was not a feature of the code received for the first experiment.

As with the first experiment submissions, no solution was submitted that was completely immune from all of the common issues identified. Over half the participants failed to ever check the sender of incoming messages. Whenever this check was performed, it was always done by means of a hard-coded agent name. For the \textit{Progress} issue, almost a quarter of subjects made some attempt to check for the correct message sequence. However, no student implemented this type of checking every time it was appropriate.

We believe that the findings of the two experiments described provide a strong argument in favour of the use of a conversation-handling technology such as ACRE that provides automated conversation checking and exception handling facilities without adding to the overall effort a programmer must go to when programming MASs in which communication plays a significant part.

\section{Comparison with Jason} \label{sec:jason}

The issues identified above arose specifically within AOP code using one specific programming language (AF-AgentSpeak) on one agent platform (Agent Factory). To show that ACRE's conversation-handling capabilities could have a wider benefit than this single configuration, it was necessary to perform further analysis. To this end, we sought to examine how conversations are handled in a different agent framework that also lacks built-in conversation and protocol management. For this to be effective, we required that some best-practice conversation handling code must be available, so that any issues identified would not be as a result of poor coding practice or a lack of familiarity with the full capabilities of the language or platform.

Jason is a MAS development platform that makes use of an extended version of AgentSpeak(L) as its AOP language~\cite{Bordini2007}. It supports inter-agent communication using KQML-like messages. However, it lacks built-in support for conversation handling and protocol definitions. Jason was chosen for this analysis for two principal reasons. Firstly, it is a popular, well-known platform. Secondly, a book is available that was written by Jason's developers that includes sample code for performing a variety of tasks, including inter-agent communication~\cite{Bordini2007}. As this code is written by the platform's developers, we assume that it represents recommended best-practice.

Figure~\ref{fig:jason-contract-net} is an extract from an implementation of a contract net protocol for Jason~\cite[p. 134]{Bordini2007}. This implementation is provided by the developers of Jason to illustrate how an interaction protocol may be implemented for that platform. The extract shows a plan that makes up part of the agent that initiates and coordinates the contract net. It is intended to be used when all bids have been received, so as to find the winner (line 7) and create an intention to announce the result to all the participants (line 9).

\begin{figure}[!ht]

	\begin{Verbatim}[numbers=left,fontsize=\footnotesize]
@lc1[atomic]
+!contract(CNPId) : cnp_state(CNPId,propose) <-
   -+cnp_state(CNPId,contract);
   .findall(offer(O,A),propose(CNPId,O)[source(A)],L);
   .print("Offers are ",L);
   L \== []; // constraint the plan execution to at least one offer
   .min(L,offer(WOf,WAg)); // sort offers, the first is the best
   .print("Winner is ",WAg," with ",WOf);
   !announce_result(CNPId,L,WAg);
   -+cnp_state(Id,finished).
\end{Verbatim}
	\caption{Sample Jason rule forming part of an implementation for  Contract Net protocol}
	\label{fig:jason-contract-net}
\end{figure}

In the same way as the trading game presented in this paper, the sample MAS in which this agent was designed to run consists of a fixed set of agents with a particular purpose. Specifically, all other agents in the system were intended as participants in the contract net. As such, no allowance is made in the code for proposals being received from agents that were not party to the initial call for proposals. This can be seen in line 4 of the extract, which creates a list of offers that have been received based on any proposal that has been received from any source. This is the same as the \textit{Sender} issue identified above. Extending this code for a more open MAS would require modification of the code to check that agents sending proposals are expected to do so. Jason does provide a method named \texttt{SocAcc} (meaning ``socially acceptable'') that can prevent some types of message being processed if they are sent by inappropriate agents. Although this could be used to prevent non-participating agents from sending proposals, it is not sufficiently fine-grained to prevent an agent that is a party to one contract net from sending a proposal relating to another.

Figure~\ref{fig:jason-contract-net} also illustrates that a Jason agent could also be susceptible to the ``Progress'' issue. The \texttt{cnp\_state} belief in this extract is used to track the state of the conversation. As the code in question is written by experts, this belief is present in a number of plans so that the state of the contract net conversation is known at all times. However, as our evaluation has shown, less experienced programmers are more prone to omitting this type of checking.

Another issue arises in the choice of an identifier for the conversation (referred to as the \texttt{CNPId} variable in the extract shown). As presented, this ID is manually specified in the original intention that triggers the start of the contract net (not shown). ACRE assigns IDs to conversations automatically, meaning that the programmer need not be concerned with this aspect of conversation handling.

From this analysis, it can be seen that in the absence of integrated conversation handling, AOP developers using Jason are also susceptible to the issues outlined above. We believe that the type of conversation handling ACRE provides would help to avoid these pitfalls and so aid the development of reliable, scalable protocol implementations.

\section{Conclusions and Future Work} \label{sec:conclusions}

We have described experiments whereby two groups of students were required to solve a communication-focused problem with and without the use of ACRE. Objective metrics applied to the first experiment indicated that ACRE can reduce the amount of code required to implement the protocols provided when compared to implementing protocols without ACRE. However, similar metrics did not produce a statistically significant difference in the second case, although it should be noted that the sample size was much smaller. The metrics used did not indicate that ACRE adds to the effort required of programmers to implement conversations, which suggests that its learning curve is not overly steep.

On further subjective analysis, a number of issues arose with non-ACRE code. These would require substantial modification of the code if the scenario were extended by the addition of additional Player agents, duplicate core agents, similar protocols or malicious agents of any type. The issues observed cannot occur with the use of ACRE, as the automatic conversation management ensures that both message senders and sequence are checked without developer intervention. These issues were present in code submitted as part of both experiments and no submission was received that was immune from all problems identified.

We also analysed some best-practice conversation-handling code written for Jason and observed that the issues identified are applicable to that platform. We therefore suggest that a conversation management framework such as ACRE is generally desirable to aid the development of communication-heavy MASs.


\subsection{Evolution of the Trading Game}

The scenario as described has potential for further refinement in order to be usable as a more general-purpose evaluation platform.

\begin{itemize}
   \item In its current guise, the trading scenario accommodates a single Player agent. A logical next step to take would be to allow a multi-player head-to-head to take place. This would mean that Player agents are in direct competition for auctions, as well as potentially creating a situation whereby one Player agent may take advantage of poorly-coded opponents by exploiting one or more of the issues identified in the above analysis.
   \item Although the facility to save and re-run particular games is possible, at present no predefined games have been created. A library of game configurations would allow for a variety of scenarios where, for example, auctions would take greater or lesser importance so a single strategy would not necessarily be best in all situations.
   \item The system agents provided as part of the trading game all behave as expected, meaning that none of them send out-of-sequence messages to test the error-handling of the Player agents. A more difficult trading game would require this to be handled.
   \item As an alternative to multi-player games, a malicious rogue system agent could be included. The behaviour of this agent would not be described in the specification other than to state that it may send any message at any time. This would attempt to impersonate the existing system agents to exploit agents that do not check messages senders. It would be interesting to measure the effect this additional requirement would have on the time taken to develop implementations of protocols.
\end{itemize}

\bibliography{ProMAS2012}

\begin{thebibliography}{10}

\bibitem{Lillis2011}
Lillis, D., Collier, R.W.:
\newblock {Augmenting Agent Platforms to Facilitate Conversation Reasoning}.
\newblock In Dastani, M., Seghrouchni, A.E.F., Hubner, J.F., Leite, J., eds.:
  Post-proceedings of the 3rd International Workshop on LAnguages,
  methodologies and Development tools for multi-agent systemS, Lyon, France,
  Springer (2011)

\bibitem{Muldoon2009}
Muldoon, C., O'Hare, G.M.P., Collier, R.W., O'Grady, M.J.:
\newblock {Towards Pervasive Intelligence : Reflections on the Evolution of the
  Agent Factory Framework}.
\newblock In {El Fallah Seghrouchni}, A., Dix, J., Dastani, M., Bordini, R.H.,
  eds.: Multi-Agent Programming: Languages, Platforms and Applications and
  Applications.
\newblock Springer US, Boston, MA (2009)  187--212

\bibitem{Russell2011}
Russell, S., Jordan, H., O'Hare, G.M.P., Collier, R.W.:
\newblock {Agent Factory : A Framework for Prototyping Logic-Based AOP
  Languages}.
\newblock In: Proceedings of the Ninth German Conference on Multi-Agent System
  Technologies (MATES 2011), Berlin, Germany (2011)

\bibitem{Bordini2007}
Bordini, R.H., H\"{u}bner, J.F., Wooldridge, M.:
\newblock {Programming multi-agent systems in AgentSpeak using Jason}.
\newblock Wiley-Interscience (2007)

\bibitem{Bellifemine2010}
Bellifemine, F., Caire, G., Trucco, T., Rimassa, G.:
\newblock {JADE Programmer's Guide (JADE 4.0)} (2010)

\bibitem{Barbuceanu1995}
Barbuceanu, M., Fox, M.S.:
\newblock {COOL: A language for describing coordination in multi agent
  systems}.
\newblock In: Proceedings of the First International Conference on Multi-Agent
  Systems (ICMAS-95). (1995)  17--24

\bibitem{Cost1998}
Cost, R.S., Finin, T., Labrou, Y., Luan, X., Peng, Y., Soboroff, I., Mayfield,
  J., Boughannam, A.:
\newblock {Jackal: a Java-based Tool for Agent Development}.
\newblock In: Working Papers of the AAAI-98 Workshop on Software Tools for
  Developing Agents, AAAI Press (1998)

\bibitem{Bradshaw1997}
Bradshaw, J.M., Dutfield, S., Benoit, P., Woolley, J.D.:
\newblock {KAoS: Toward an industrial-strength open agent architecture}.
\newblock Software Agents (1997)  375--418

\bibitem{Huget2003}
Huget, M.P., Koning, J.L.:
\newblock {Interaction Protocol Engineering}.
\newblock Communications (2003)  291--309

\bibitem{Ancona2012}
Ancona, D., Drossopoulou, S., Mascardi, V.:
\newblock {Automatic Generation of Self-Monitoring MASs from Multiparty Global
  Session Types in Jason}.
\newblock In: Proceedings of Declarative Agent Languages and Technologies (DALT
  2012), Valencia, Spain (2012)  1--17

\bibitem{Hochstein2008}
Hochstein, L., Basili, V.R., Vishkin, U., Gilbert, J.:
\newblock {A pilot study to compare programming effort for two parallel
  programming models}.
\newblock Journal of Systems and Software \textbf{81}(11) (2008)  1920--1930

\bibitem{Luff2009}
Luff, M.:
\newblock {Empirically Investigating Parallel Programming Paradigms : A Null
  Result}.
\newblock In: Workshop on Evaluation and Usability of Programming Languages and
  Tools (PLATEAU). (2009)  43--49

\bibitem{Rossbach2010}
Rossbach, C.J., Hofmann, O.S., Witchel, E.:
\newblock {Is Transactional Programming Actually Easier?}
\newblock ACM SIGPLAN Notices \textbf{45}(5) (2010)  47--56

\bibitem{VanderWiel1997}
VanderWiel, S.P., Nathanson, D., Lilja, D.J.:
\newblock {Complexity and performance in parallel programming languages}.
\newblock In: Second International Workshop on High-Level Programming Models
  and Supportive Environments. (1997)  3--12

\bibitem{Rao1996}
Rao, A.S.:
\newblock {AgentSpeak (L): BDI agents speak out in a logical computable
  language}.
\newblock Agents Breaking Away (1996)

\end{thebibliography}

\end{document}